\documentclass[prb,aps,twocolumn,groupedaddress]{revtex4}

\usepackage{graphicx}

\usepackage{amssymb}

\begin{document}


\title{Resistive switching effects on the spatial distribution of phases in metal-complex oxide interfaces}

\author{A. Schulman}
\affiliation{Laboratorio de Bajas Temperaturas, Departamento de
F\'{\i}sica, FCEyN, UBA, and IFIBA (CONICET), Ciudad Universitaria,
(C1428EHA) Buenos Aires, Argentina}
\author{C. Acha}
\thanks{Corresponding author: acha@df.uba.ar}
\affiliation{Laboratorio de Bajas Temperaturas, Departamento de
F\'{\i}sica, FCEyN, UBA, and IFIBA (CONICET), Ciudad Universitaria,
(C1428EHA) Buenos Aires, Argentina}

\begin{abstract}
In order to determine the key parameters that control the resistive
switching mechanism in metal-complex oxides interfaces, we have
studied the electrical properties of metal /
YBa$_2$Cu$_3$O$_7-\delta$ (YBCO) interfaces using metals with
different oxidation energy and work function (Au, Pt, Ag) deposited
by sputtering on the surface of a YBCO ceramic sample. By analyzing
the IV characteristics of the contact interfaces and the temperature
dependence of their resistance, we inferred that ion migration may
generate or cancel conducting filaments, which modify the resistance
near the interface, in accordance with the predictions of a recent
model.
\end{abstract}

\maketitle

\section{Introduction}
\label{Intro}

The resistive random access memories (RRAM) based on the
non-volatile change of the resistance of metal-oxide interfaces upon
the application of an electric field are nowadays an interesting
emerging technology.~\cite{Freitas08} Their possible technological
application as a replacement of the actual non-volatile RAM memory
devices is encouraging new basic research in order to obtain a
perfect control of all the parameters that influence the effect.
Although many papers have been published in the last years on the
resistive switching (RS) effect, observed in many metal-oxide
interfaces,~\cite{beck,liu,tulina,seo,fors,Waser07,Quintero05,Sawa08,Acha09a,Acha11a}
many aspects of the mechanism are still an open question. However,
in a recent paper, Rozenberg et al.\cite{Rozenberg10} introduce a
model that accounts for the bipolar RS phenomenon observed in
transition metal oxides, highlighting the key role played by oxygen
vacancies in this mechanism. Here, we analyze the temperature
dependence of the RS effect in metal / ceramic YBCO interfaces with
the objective to gain insight on the relevance of thermal energy on
the particular characteristics of the RS change. Our results
indicate that the observed temperature dependence of the resistance
shift during the RS is essentially associated with the temperature
dependence of the resistance of each "on" and "off" states.

\section{Experimental}
\label{Expe}

\begin{figure} [t]
\centerline{\includegraphics[angle=0,scale=0.28]{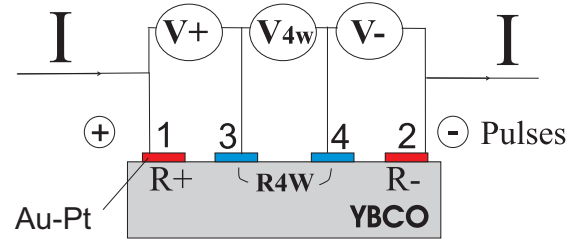}}
\vspace{-2mm} \caption{(Color online) Contact configuration used to
study the RS of YBCO / metal interfaces.} \vspace{0mm}
\label{fig:contactos}
\end{figure}

To study the RS effect in metal / YBCO interfaces we sputtered 4 Au
(Pt or Ag) electrodes (labeled 1,2,3 and 4) on one of the faces of a
ceramic YBCO as it is depicted in Fig.~\ref{fig:contactos}.
Characteristics of the textured ceramic YBCO sample can be found
elsewhere.\cite{Acha09a}

Our pulsing protocol (on electrodes 1 and 2) consisted of trains of
20,000 square pulses of amplitude $V_{-pulse}$ and 0.1 ms width at 1
kHz. As the pulsing generates a maximum excursion of $\sim$ 1 K on a
thermometer well thermally anchored to the sample during a transient
of $\sim$ 20 s, we wait 60 s to ensure a stable temperature
condition during the resistance measurement. To measure the remanent
resistance after the pulsing treatment, a small bias current is
applied to electrodes 1 and 2: by measuring the voltage drop in
electrodes 1 and 3 we essentially evaluate the resistance near the
interface corresponding to electrode 1 as the resistance in series
coming from the bulk part of the YBCO between electrodes is
negligible (confirmed by measuring the four terminal resistance
$R_{4W}$). The same occurs when we measure the voltage between
electrodes 4 and 2; we essentially evaluate the resistance near the
interface of electrode 2. We arbitrarily call $R_+$ the former
resistance ($V_{13}/I_{12}$) as it follows the polarity of the
applied pulses while the latter corresponds to $R_-$
($V_{42}/I_{12}$). If each remanent resistance $R_+$ or $R_-$ is
measured after applying $V_{-pulse}$ and $V_{-pulse}$ is varied
cyclically by a defined step in $\pm V_{max}$ interval, a curve
called resistance hysteresis switching loops (RHSL) can be obtained.
This curve gives a clear view of the behavior of the non-volatile
resistance upon the applied excitation of the pulses. Two terminals
IV characteristics (1-2 electrodes) were also measured, applying a
positive triangular waveform (+10 V, 6 Hz) to electrodes 1 and 2 in
series with a calibrated 100 $\Omega$ resistance, used to determine
the circulating current by measuring the voltage drop between its
terminals.

\section{Results and Discussion}
\label{R&D}

The RS effect was observed for all the metallic electrodes we used
(Au, Pt, Ag), independently of their oxidation energy or their work
function. The main difference observed for these metal-YBCO
interfaces is associated with the low resistance state, where higher
values were measured with decreasing the oxidation energy of the
metal. Additionally, Pt-YBCO interfaces gave RHSL of lower amplitude
than Au-YBCO and Ag-YBCO, while the latter shows poor endurance.
Here, we mainly show the results obtained for the Pt electrodes,
which can be considered representative of the metallic interfaces
studied in this work, as similar temperature effects on RS
characteristics were obtained for Au and Ag electrodes.

\begin{figure} [ht]
\centerline{\includegraphics[angle=0,scale=0.3]{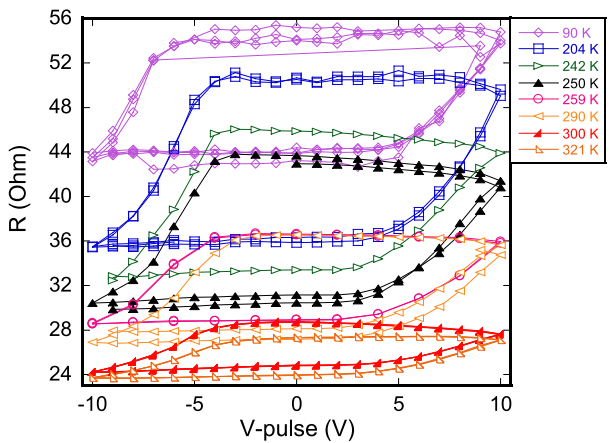}}
\vspace{-2mm} \caption{(Color online) Resistance hysteresis
switching loops (RHSL) at different temperatures corresponding to
the non-volatile value of $R_+$ after applying a pulsing protocol of
amplitude $V_{-pulse}$ (Pt-YBCO interfaces).} \vspace{0mm}
\label{fig:RHSLvsT}
\end{figure}

The RHSL curves of $R_+$ at different temperatures can be observed
in Fig.\ref{fig:RHSLvsT}. The results for $R_-$, not shown here, are
similar but show a complementary behavior, as when $R_+$ switches to
its low or "on" value, $R_-$ do it to its high or "off"
state.\cite{Acha09b} Decreasing the temperature clearly produces an
increase of the resistance change between the two states.

\begin{figure} [ht]
\centerline{\includegraphics[angle=0,scale=0.35]{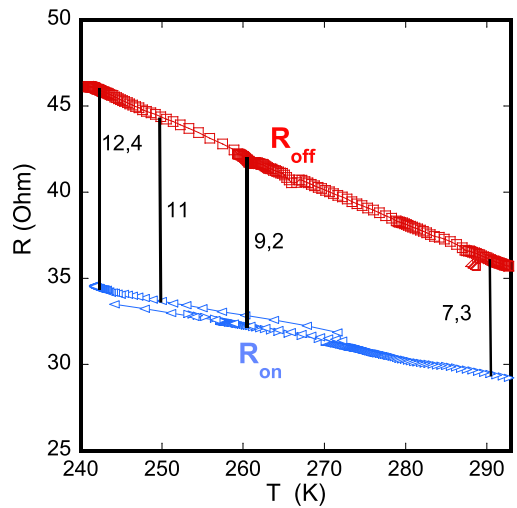}}
\vspace{-2mm} \caption{(Color online) Temperature dependence of the
$R_+$ electrode in its "on" and "off" states (Pt-YBCO interfaces).
The lines indicate the shift of resistance produced by a RS at each
particular temperature.} \vspace{0mm} \label{fig:RvsT}
\end{figure}

\begin{figure} [hb]
\centerline{\includegraphics[angle=0,scale=0.29]{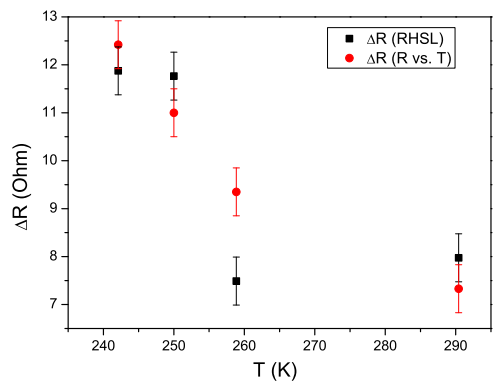}}
\vspace{-2mm} \caption{(Color online) Comparison between the
temperature dependent resistance difference between the "on" and
"off" states and the resistance shift produced at each temperature
in a RHSL (Pt-YBCO interfaces).} \vspace{0mm} \label{fig:saltos}
\end{figure}

The temperature dependence of $R_+$ in its "on" and "off" states
(switched at room temperature) is shown in Fig.\ref{fig:RvsT}. For
both states a semiconducting-like dependence can be observed. As a
consequence, their resistance difference increases with decreasing
temperature. Interestingly, this temperature dependent difference is
very similar to the resistance change obtained during each RHSL, as
it is indicated by the lines in Fig.\ref{fig:RvsT} and can also be
observed in Fig.\ref{fig:saltos}.

We have previously shown\cite{Acha09a} that pulsing can suppress and
restore the superconducting state of the YBCO material in the
neighborhood of the pulsed electrodes, by creating or destroying
YBCO filaments, affecting the geometrical conducting factor near the
interfaces. This also occurs with other oxides, where the contact
resistance "copies" the temperature dependence of the bulk
oxide\cite{Quintero05}, probably due to the fact that the conduction
near the interface is also related to conducting filaments immersed
in an insulating matrix.

Within this framework, the room temperature switching between the
"off" and the "on" states (or viceversa) may be produced by the
migration of oxygen ions that creates (or destroys) a set of well
conducting YBCO filaments.\cite{Rozenberg10} The coincidence of the
resistance shift during the RHSL at each temperature with the
resistance difference between the "on" and "off" states indicates
that it is essentially the same (or an equivalent) set of filaments
that is involved in the RS, independently of the temperature. In
that sense, the electric field induced-spatial distribution of the
conducting and insulating phases remains unchanged by varying the
temperature at which the pulsing treatment is applied.

\begin{figure} [ht]
\centerline{\includegraphics[width=7cm,height=6cm]{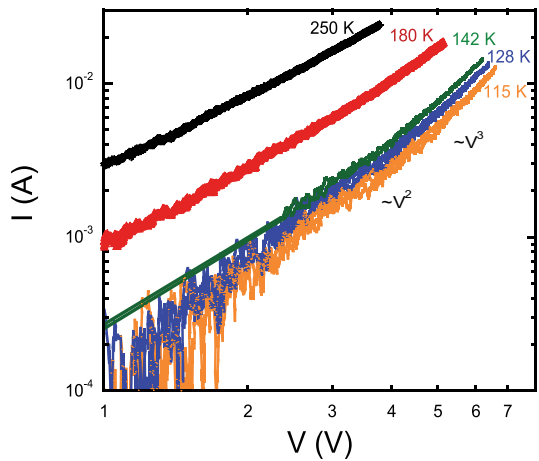}}
\vspace{-2mm} \caption{(Color online) Two terminals IV
characteristics at different temperatures (Au-YBCO interfaces). A
power law $I \sim V^n$ with an increasing $n$ with increasing V is
characteristic of a SCLC conduction mechanism.} \vspace{0mm}
\label{fig:IVvsT}
\end{figure}

The temperature dependence of the IV characteristics of electrodes
1-2 in series, shown in Fig.\ref{fig:IVvsT}, are consistent with
this scenario. The conduction mechanism at low temperatures is of
the space charge limited currents (SCLC) type.\cite{Dietz95} This
mechanism occurs when the conduction interface is formed by a
metallic layer, capable of injecting carriers, and a semiconductor
in close contact. The distribution of well oxygenated YBCO
conducting zones, structurally connected to YBCO oxygen depleted
surroundings may account for the observed characteristics.

\section{Conclusions}
\label{Conc} We have studied the RS of different metal (Au, Pt, Ag)
-YBCO interfaces obtaining similar effects independently of the
metal used. We have shown that the temperature dependence of the
resistance shift in the RHSL coincides with the difference of the
resistances between the "on" and "off" states obtained during a RS
at room temperature. This result indicates that although the thermal
energy modifies the oxygen diffusion responsible of the RS, the main
effect of temperature is associated with the temperature dependence
of the resistance of a set of conducting filaments near the
interface, which can be broken or restored by the pulsing treatment,
regardless of the temperature at which the process was conducted.

\section{Acknowledgments}
\label{Acknow} This work was supported by CONICET Grant PIP
112-200801-00930 and UBACyT X166 (2008-2010). We acknowledge
fruitful discussions with P. Levy, M. J. Rozenberg, M. J. S\'anchez
and R. Weht. We are indebted to D. Gim\'enez, E. P\'erez Wodtke and
D. Rodr\'{\i}guez Melgarejo for their technical assistance.




\end{document}